\documentclass{nature}
\usepackage{amsmath, amsthm, amssymb, mathrsfs, graphicx, multibib}
\usepackage{color,ulem}
\newcites{method}{Methods References}%
\newcommand{\ujybeam}{$\mu$Jy beam$^{-1}\,$}
\renewcommand{\deg}{$^o\,$}
\newcommand{\A}{{\it A}}
\newcommand{\B}{{\it B}}
\newcommand{\C}{{\it C}}
\newcommand{\D}{{\it D}}
\newcommand{\msun}{\,{\rm M_{\odot}}}
\newcommand{\cm}{\,{\rm cm}}

\newcommand{\erg}{\,{\rm erg}}
\newcommand{\s}{\,{\rm s}}
\definecolor{green}{rgb}{0.0, 0.5, 0.0}
\title{Superluminal motion of a relativistic jet in the neutron star merger GW170817}

\begin{document}

\maketitle

% % Uncomment for arxiv submission
\noindent K. P. Mooley (1,2,10,*), A. T. Deller (3,4,*), O. Gottlieb (5,*), E. Nakar (5), G. Hallinan (2), S. Bourke (6), D.A. Frail (1), A. Horesh (7), A. Corsi (8), K. Hotokezaka (9) (Affiliations: (1) NRAO, (2) Caltech, (3) Swinburne University of Technology, (4) OzGrav, (5) Tel Aviv University, (6) Chalmers University of Technology, (7) Hebrew University of Jerusalem, (8) Texas Tech University, (9) Princeton University, (10) Jansky Fellow). $^*$ These authors contributed equally to this work
\vspace{1cm}

% \section*{Authors and Affiliations}
% comment out for arxiv submission
% \noindent
% \author{K. P. Mooley$^{1,2,10,*}$, A. T. Deller$^{3,4,*}$, O. Gottlieb$^{5,*}$, E. Nakar$^{5}$, G. Hallinan$^{2}$, S. Bourke$^{6}$, D.A. Frail$^{1}$, A. Horesh$^{7}$, A. Corsi$^{8}$ \& K. Hotokezaka$^{9}$}

% comment out for arxiv submission
% \begin{affiliations}
% \item National Radio Astronomy Observatory, Socorro, New Mexico, 87801, USA
% \item Caltech, 1200 E California Blvd, MC 249-17, Pasadena, CA 91125, USA
% \item Centre for Astrophysics \& Supercomputing, Swinburne University of Technology, John St, Hawthorn VIC 3122 Australia
% \item ARC Centre of Excellence for Gravitational Wave Discovery (OzGrav), Australia
% \item The Raymond and Beverly Sackler School of Physics and Astronomy, Tel Aviv University, Tel Aviv 69978, Israel
% \item Department of Space, Earth and Environment, Chalmers University of Technology, Onsala Space Observatory, S-439 92 Onsala, Sweden
% \item Racah Institute of Physics, The Hebrew University of Jerusalem, Jerusalem, 91904, Israel
% \item Department of Physics and Astronomy, Texas Tech University, Box 41051, Lubbock, TX 79409-1051, USA
% \item Department Astrophysical Sciences, Princeton University, Peyton Hall, Princeton, NJ 08544, USA
% \item Jansky Fellow (NRAO, Caltech)
% \end{affiliations}
% \vspace{-1cm}
% \noindent $^*$ These authors contributed equally to this work

% \clearpage
\begin{abstract}
% The binary neutron star merger GW170817 was accompanied by radiation across the electromagnetic spectrum and localized to the galaxy NGC 4993 at a distance of 41+/-3 Mpc. The radio and X-ray afterglows of GW170817 exhibited delayed onset, a gradual rise in the emission with time as t^0.8, a peak at about 150 days post-merger, followed by a relatively rapid decline. To date, various models have been proposed to explain the afterglow emission, including a choked-jet cocoon and a successful-jet cocoon (a.k.a. structured jet). However, the observational data have remained inconclusive as to whether GW170817 launched a successful relativistic jet. Here we show, through Very Long Baseline Interferometry, that the compact radio source associated with GW170817 exhibits superluminal motion between two epochs at 75 and 230 days post-merger. This measurement breaks the degeneracy between the models and indicates that, while the early-time radio emission was powered by a wider-angle outflow (cocoon), the late-time emission was most likely dominated by an energetic and narrowly-collimated jet, with an opening angle of <5 degrees, and observed from a viewing angle of about 20 degrees. The imaging of a collimated relativistic outflow emerging from GW170817 adds substantial weight to the growing evidence linking and short gamma-ray bursts. 

The binary neutron star merger GW170817\cite{abbott2017a} was accompanied by radiation across the electromagnetic spectrum\cite{abbott2017b} and localized\cite{abbott2017b} to the galaxy NGC 4993 at a distance\cite{hjorth2017} of 41$\pm$3 Mpc.
The radio and X-ray afterglows of GW170817 exhibited delayed onset\cite{hallinan2017,troja2017,margutti2017,haggard2017}, a gradual rise\cite{mooley2018} in the emission with time as $t^{0.8}$, a peak at about 150 days post-merger\cite{dobie2018}, followed by a relatively rapid decline\cite{dobie2018,alexander2018}. 
To date, various models have been proposed to explain the afterglow emission, including a choked-jet cocoon\cite{kasliwal2017,hallinan2017,mooley2018,troja2018,xie2018} and a successful-jet cocoon\cite{kasliwal2017,hallinan2017,lamb2017,mooley2018,troja2018,lazzati2018,margutti2018,Lyman2018,xie2018,resmi2018} (a.k.a. structured jet). However, the observational data have remained inconclusive\cite{nakar-piran2018,lazzati2018,troja2018-atel,alexander2018} as to whether GW170817 launched a successful relativistic jet. 
Here we show, through Very Long Baseline Interferometry, that the compact radio source associated with GW170817 exhibits superluminal motion between two epochs at 75 and 230 days post-merger. This measurement breaks the degeneracy between the models and indicates that, while the early-time radio emission was powered by a wider-angle outflow\cite{mooley2018} (cocoon), the late-time emission was most likely dominated by an energetic and narrowly-collimated jet, with an opening angle of $<5$ degrees, and observed from a viewing angle of about 20 degrees. The imaging of a collimated relativistic outflow emerging from GW170817 adds substantial weight to the growing evidence
%\cite{eichler1989,nakar2007,perley2009,tanvir2013,berger2013,berger2014} 
linking binary neutron star mergers and short gamma-ray bursts. 

\end{abstract}

\begin{figure}
\centering
\includegraphics[width=6in]{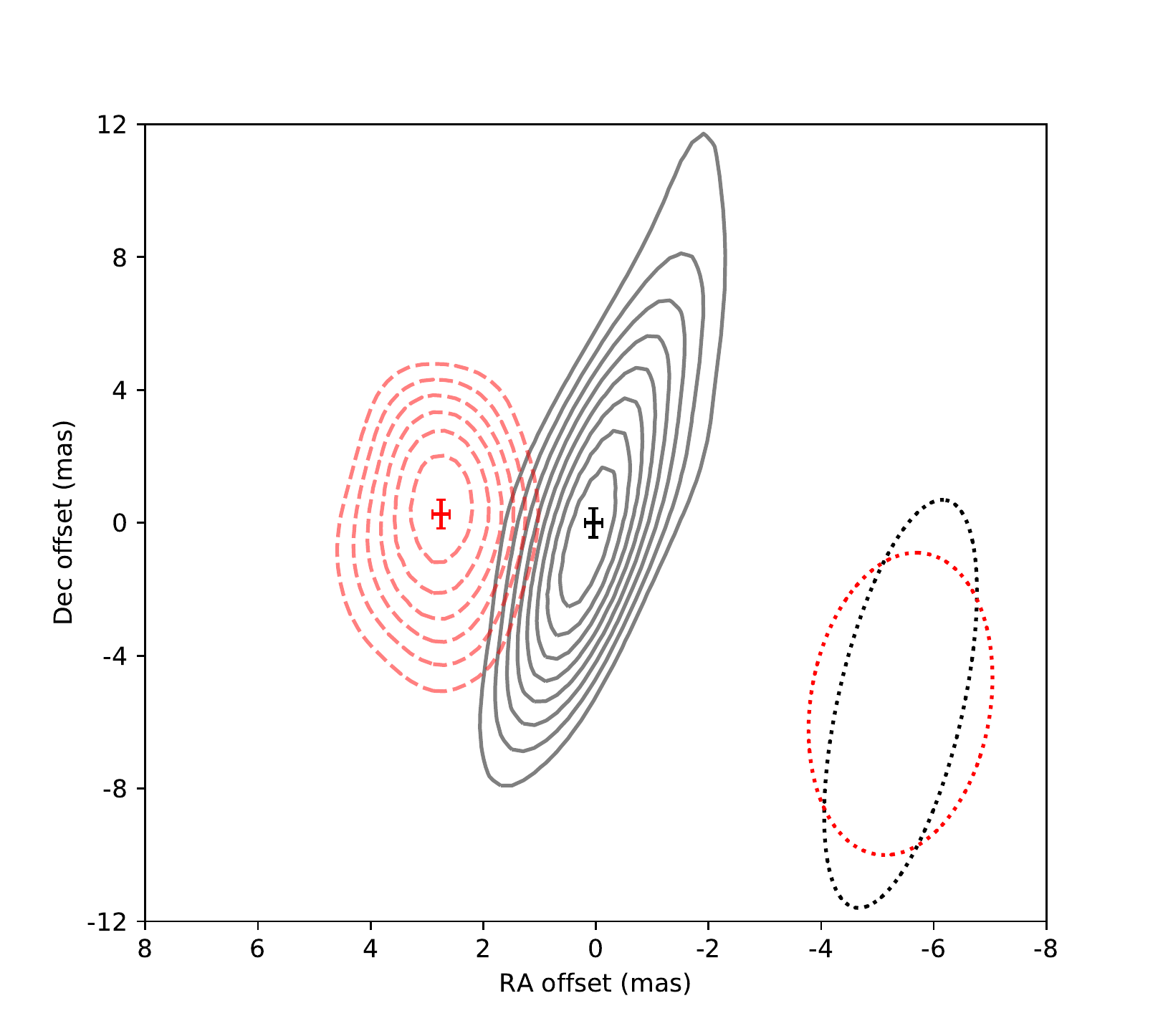}
\caption{{\bf Proper motion of the radio counterpart of GW170817.}
The centroid offset positions (shown by 1$\sigma$ errorbars) and 3$\sigma$-12$\sigma$ contours of the radio source detected 75 d (black) and 230 d (red) post-merger with Very Long Baseline Interferometry (VLBI) at 4.5 GHz. The two VLBI epochs have image RMS noise of 5.0 \ujybeam and 5.6 \ujybeam (natural-weighting) respectively, and the peak flux densities of GW170817 are 58 \ujybeam and 48 \ujybeam respectively. The radio source is consistent with being unresolved at both epochs. The shape of the synthesized beam for the images from both epochs are shown as dotted ellipses to the lower right corner. The proper motion vector of the radio source has a magnitude of $2.7\pm0.3$ mas and a position angle of $86^o\pm18^o$, over 155 d.}
\label{fig:proper_motion}
\end{figure}

Our VLBI observations with the High Sensitivity Array (HSA), comprising of the Very Long Baseline Array (VLBA), the Karl G. Jansky Very Large Array (VLA) and the Robert C. Byrd Green Bank Telescope (GBT), 75 d and 230 d post-merger (mean epochs; see Methods), indicate that the centroid position of the radio counterpart of GW170817 changed from RA$=$13:09:48.068638(8), Dec=$-$23:22:53.3909(4) to RA=13:09:48.068831(11), Dec=$-$23:22:53.3907(4) between these epochs (brackets quote 1$\sigma$ uncertainties in the last digits). This implies an offset of 2.67$\pm$0.19$\pm$0.21 mas in RA and 0.2$\pm$0.6$\pm$0.7 mas in Dec (1$\sigma$ uncertainties, statistical and systematic respectively; see Methods). This corresponds to a mean apparent velocity of the source along the plane of the sky $\beta_{\rm app}=4.1\pm0.5$, where $\beta_{\rm app}$ is in units of the speed of light, c (1$\sigma$, including the uncertainty in the source distance). Offset positions of the radio source and the positional uncertainties at both VLBI epochs are shown in Figure~\ref{fig:proper_motion}. Our VLBI data are consistent with the radio source being unresolved both at day 75 and day 230. Given the VLBI angular resolution and the signal--to--noise ratio of the detection, this puts an upper limit on the source size in both epochs of about 1 mas (0.2 pc at the distance of NGC 4993) in the direction parallel to the source motion and 10 mas perpendicular to the source motion (see Methods). 

% However, unlike GW170817, the size of the radio source associated with GRB 030329 was seen to increase faster than the speed of light, and it did not show any detectable motion of the centroid. This indicated that the relativistic outflow GRB 030329 pointed towards us. As we discuss below, this is not the case for GW170817.

The significant proper motion of the radio source immediately rules out isotropic ejecta models\cite{hotokezaka2018-afterglow,davanzo2018,gill-granot2018} for the radio (and X-ray) afterglow, which predict proper motion close to zero, and argues in favor of highly anisotropic ejecta (consistent with jet models). If the ejecta are bipolar, then one of the components is relativistically beamed into our line of sight. 

While superluminal motion is seen frequently in active galactic nuclei and micro-quasars, it is extremely rare in extragalactic explosive transients. Superluminal motion has been measured only in one such transient: the long-duration GRB 030329\cite{taylor2004}. GRB 030329 had a measured superluminal expansion ($\beta_{\rm app}\approx3-5$) but no proper motion, while GW170817 has measured proper motion but no expansion. While both were relativistic events of comparable energies, these differences immediately suggest different geometries and/or viewing angles. 

% \section*{Analytical constraints on the geometry and source size}
% The two VLBI epochs were observed before and slightly after the peak of the light curve (about day 150). The light curve turnover after the peak is very rapid, where our most recent flux density measurement shows that by day 240 the flux falls roughly as $F_\nu \propto t^{-2}$ (Mooley et al. in prep). The rising light curve and the turnover at day 150 could be explained by a range of outflow structures, ranging from narrowly collimated to spherical (see above). As we show below, the superluminal motion of the image enables us to break the degeneracy and determine the outflow structure at a rather high confidence. When combined with the rapid decay in the light curve, we can put tight constraints on the entire geometry of the event, including the opening angle of the emitting region and the viewing angle.

The apparent velocity and size of a source moving at relativistic speeds, such as the radio counterpart of GW170817, differs from the actual velocity and size. The image of a point source, for example, moving at a Lorentz factor $\Gamma$ and viewed at an angle $\theta$, is point-like and has a maximal apparent velocity of $\beta_{app}=\Gamma$, which is obtained when $\theta=1/\Gamma$. On the other hand, the maximal centroid velocity of an extended source with a uniform $\Gamma$ is smaller than $\Gamma$, and its image size increases\cite{boutelier2011} with the source size and with $\Gamma$. An extreme example of the latter case is a spherically symmetric source expanding isotropically. In such a case, the image is a ring with a radius that increases at a velocity $\Gamma$ with no centroid motion. The centroid velocity may also be affected in cases where we see different regions of the outflow at different times\cite{lind1985} (i.e. a pattern motion).

Using this information, we now examine the results from the VLBI data and the radio light curve to derive analytical constraints on the geometry and source size. We assume that the ejecta is axis-symmetric, such that $\theta_{\rm obs}$ is the viewing angle and $\theta_{\rm s}$ is the average angular size of the source that dominates the emission between days 75 and 230 days post-merger (both with respect to the symmetry axis). 
If the source is compact ($\theta_{\rm s} \lesssim \theta_{\rm obs}-\theta_{\rm s})$, then the source size and possible pattern motion has minor effects and we can use the point source approximation. In all the highly aspherical models suggested, the energy density increases towards the axis of symmetry, implying that during the peak of the light curve the emission is dominated by a region at $(\theta_{\rm obs}-\theta_{\rm s}) \sim 1/\Gamma$. Using the point source approximation this implies that between the two observations the source is observed at an angle $(\theta_{\rm obs}-\theta_{\rm s}) \approx 1/\beta_{\rm app} \approx 0.25$ rad and its Lorentz factor is $\Gamma \approx \beta_{\rm app} \approx 4$. If the source is extended $(\theta_{\rm s} \gg \theta_{\rm obs}-\theta_{\rm s})$, then in order to achieve the observed apparent velocity the source should have $\Gamma>4$ and possibly $\theta_{\rm obs}-\theta_{\rm s}<0.25$ rad. 

%While the VLBI measurement cannot rule out a somewhat extended source (especially in the direction perpendicular to the proper motion),
There are several strong lines of evidence suggesting that the source is compact.  First, the source is very compact in our VLBI observations, and is consistent with being unresolved. Second, the observed flux depends very strongly on $\Gamma$ (roughly as $\Gamma^{10.4}$), implying that on day 150 the Lorentz factor of the radio source is\cite{nakar-piran2018} $\Gamma\lesssim5$. Finally, and most constraining, is the rapid turnover around the peak of the radio light curve and the very fast decline that follows $F_\nu \propto t^{-2}$ after day 200 (K.P.M. et al., in preparation). The shape of the peak and the following decline depends on the ratio $\frac{\theta_{\rm s}}{\theta_{\rm obs}-\theta_{\rm s}}$. A smaller ratio results in a narrower peak and if $\theta_{\rm s}\gg \theta_{\rm obs}-\theta_{\rm s}$ the decay is expected to be\cite{nakar-piran2018} at first roughly linear in time, while if $\theta_{\rm s}\ll \theta_{\rm obs}-\theta_{\rm s}$ the flux decay after the peak is predicted to be roughly as $F_\nu \propto t^{-p}$, where the radio spectrum dictates\cite{margutti2018,troja2018,mooley2018} $p\approx2.16$. We conclude that the combination of the image and the light curve indicate that around the peak, at 150 d, the emission is most likely dominated by a narrow component with $\theta_{\rm s} \ll 0.25$ rad and $\Gamma\approx 4$ which is observed at an angle $\theta_{\rm obs}-\theta_{\rm s} \approx 0.25$ rad (this is in contrast to the emission during the first month or two which was most likely dominated by cocoon emission from larger angles than $\theta_{\rm s}$).

The constraints derived above strongly disfavor an uncollimated choked jet, where the jet has a wide opening angle and does not successfully escape the neutron-rich material dynamically ejected during the merger (i.e. it is choked, and hence does not contain a relativistic narrow core). A narrowly collimated choked jet may generate an outflow with a narrow high-energy core, but it is hard to obtain a Lorentz factor that is high enough without a fine tuning of the location where the jet is choked. In contrast to all other models, the successful jet model predicts a structure that can easily satisfy the constraints of the image and the light curve. In this model, the gradual rise is generated by cocoon emission and the peak is observed when the core of the successful jet decelerates and starts dominating the emission. The jet opening angle, $\theta_{\rm j}$, and its Lorentz factor are those of the source in our images around the time of the peak, namely $\theta_{\rm j} \approx \theta_{\rm s}$. We can only put a lower limit on the initial Lorentz factor of the jet, $\Gamma_0$, since we do not know the deceleration radius (i.e. when the transition from the coasting phase to the power-law decline phase took place). All the observational data can be explained with a narrowly-collimated jet having $\Gamma_0\gtrsim10$.

% \section*{Numerical simulations}
In order to verify the analytical considerations discussed above, and to find tighter constraints on the outflow, we ran a set of relativistic hydrodynamic simulations (see Methods). 
% where we inject a relativistic jet into a sub-relativistic ejecta and follow the interaction of the resulting outflow with the external medium. We then calculate the light curve and image produced by this interaction by post-processing of the hydrodynamic results. For each simulation we put the observer at different viewing angles and chose the external density and microphysical parameters so the peak of the light curve fits the observed one. Then tested whether the rise and fall of the light curve fits that data and whether the images fit the VLBI observations. 
Our simulations include configurations of choked and successful jets at various opening angles and various viewing angles, and include emission from all components of the outflow. Figure~\ref{fig:light_curves} shows light curves from six different configurations, and Figure~\ref{fig:images} shows the corresponding images at day 75 and day 230. 
As expected, we find that in the simulations where the jet is choked, the centroid velocity of the images is too slow to explain the proper motion of GW170817 and the decline of the light curve after the peak is much slower than t$^{-2}$. Among the successful jet simulations, those that were observed from a large angle, $\theta_{\rm obs}-\theta_{\rm j} \gtrsim 0.4$ rad, did not produce images that moved fast enough, while the images of jets that were observed at an angle that is too small, $\theta_{\rm obs}-\theta_{\rm j} \lesssim 0.2$ rad, the image centroid moved too fast and/or the source size was too large. The light curve also constrained the geometry and only simulations with $\frac{\theta_{\rm j}}{\theta_{\rm obs}-\theta_{\rm j}}$ that is small enough can fit the rapid transition from a rising light curve to the observed decay. Among all the configurations we examined, only extremely narrow jets with $\theta_{\rm j}<0.1$ rad that were observed at an angle of $0.2 < \theta_{\rm obs}-\theta_{\rm j} < 0.4$ rad result in emission that is consistent with the light curve and that reproduces the observed motion of the image centroid. Taken together, this implies that we see a narrow jet with $\theta_{\rm j}<0.1$ rad ($<$5\deg) from a viewing angle that is in the range $0.25<\theta_{obs}<0.50$ rad (14\deg--28\deg). This can be seen, for example, in Figures~\ref{fig:light_curves} and \ref{fig:images} where the centroid motion for models with viewing angles outside of this range deviate significantly (by more than 2$\sigma$; see Methods) from the observations and models with wider jets where $\theta_{\rm j} > 0.1$ rad do not reproduce the rapid decay after the light curve peak.
In a different study\cite{hotokezaka2018-h0}, we have carried out a full scan of the parameter space using two different semi-analytical jet structures and the values obtained for $\theta_{\rm j}$ and $\theta_{\rm obs}$ lie within the range specified above.

Our simulation that provides the best fit to the data is of a 0.08 rad (4\deg at the time of light curve peak) jet that is observed from $\theta_{\rm obs}=0.35$ rad (20\deg). 
%\udi{In this simulation a relativistic jet is injected into the sub-relativistic merger ejecta. The jet is followed during its propagation through the ejecta, the formation of the cocoon and the breakout of the jet and the cocoon from the sub-raltivistic ejecta. The simulation then continues to follow the interaction of the outflow (jet+cocoon) with the ISM.  When this interaction starts, the jet opening angle is 0.04 rad. The cocoon dominates the observed radio emission during the first $\sim$60 days, after which time the jet dominates. The jet expands sideways slowly during its interaction with the ISM, reaching an opening angle of 0.08 rad after $\sim$150 days at the light curve peak. }
In this simulation, the cocoon dominates the observed radio emission until about day 60, after which time the jet dominates (see Figure~\ref{fig:light_curves} and Methods). 
The Lorentz factor of the observed region drops slowly from $\Gamma\approx4$ on day 75 to $\Gamma\approx3$ on day 230. 
%We highlight that these constraints apply only to the time of the measurements. 
%The jet initial Lorentz factor was certainly higher than 4 (although we cannot say by how much) and its initial opening angle may have been much narrower than 0.1 rad (see Methods). \ore{The last last sentence is confusing, it's unclear whether it refers to simulations or not. I guess it doesn't, so I wouldn't put it here. Also what part in methods talks about its opening angle should be narrower?}
Within the framework of standard afterglow theory from a successful jet, the observations put tight constraints on additional properties of the jet and surrounding environment (see Methods). The total energy of the relativistic ejecta (jet+cocoon) is in the range $E \sim 10^{49}-10^{50}$ erg, and the external density is $n \sim 10^{-4}-5\times10^{-3}$ cm$^{-3}$. 

Our final model is qualitatively similar to jet+cocoon (also referred to as structured jet) models suggested previously\cite{lazzati2018,margutti2018,nakar2018,xie2018}. However, owing to the VLBI data as well as more up-to-date light curves, our constraints on jet opening angle and viewing angle are much tighter than previous models, and in tension with some. 
%Our constraint on the viewing angle, derived independently from simple geometric considerations, are more robust, and lie towards the lower bounds of previous estimates modeling of the afterglow and kilonova light curves \cite{parego2017,Lyman2018,margutti2018,xie2018,lazzati2018,nakar2018,resmi2018,davanzo2018,gill-granot2018,troja2018} and from gravitational waves\cite{abbott2017a,lvc2018,finstad2018}. 
The small viewing angle ($\sim$20\deg) for GW170817 is expected only in about 5\% of the mergers (not accounting for the gravitational wave polarization bias). Our best fit model suggests we were relatively lucky since the afterglow of this event as observed at larger angles would be much fainter. In our best fit numerical model, the radio emission should be detectable at a viewing angle of $\sim$30\deg, but probably too faint for detection at an angle of $\sim$40\deg. The detectability of future GW170817-like events depends on the circum-merger density. Taking our best fit model for GW170817, but increasing the density to 0.01 cm$^{-3}$ (the median density\cite{fong2015} for SGRBs; while keeping the all other values constant) we find an afterglow that is brighter by about an order of magnitude at the peak compare to that of GW170817. Such an afterglow could have been detected at a distance of 40 Mpc also at a larger viewing angle of $\sim 50^o$.

%The results of more comprehensive afterglow modeling suggest that GW170817 was a rather unusual event compared to the larger SGRB sample\cite{fong2015}. 
Our VLBI result implies that binary neutron star mergers launch relativistic narrowly collimated jets that successfully penetrate the dynamical ejecta, which is a prerequisite for the production of SGRBs (which require $\Gamma_0\gtrsim100$).
%SGRBs are highly efficient in producing gamma-rays and a typical SGRB lasts for a fraction of a second.  %the upper limit on the jet opening angle of 5\deg implies that its peak luminosity (isotropic equivalent) was $\gtrsim 3\times 10^{51}$ erg s$^{-1}$.
If GW170817 produced an SGRB pointing away from us, then its peak isotropic equivalent luminosity in gamma-rays, $L_{iso}$, was $\sim10^{52}$ erg s$^{-1}$ when observed within the jet cone, assuming that the initial opening angle of the jet was $\sim0.05$ rad.  
%Studies of the luminosity function of SGRBs (ref\cite{wanderman-piran2015} and references therein) find that the local rate of SGRBs that points towards Earth is $R_{\rm GRB} \sim 10$ Gpc$^{-3}$\,yr$^{-1}$, but this rate is dominated by SGRBs with a peak $L_{iso}$ of $\sim5\times10^{49}$ erg s$^{-1}$ (see however ref\cite{ghirlanda2016}). The rate decreases roughly linearly with the energy, and 
The rate of SGRBs with a peak $L_{iso} \gtrsim10^{52}$ erg s$^{-1}$ is only\cite{wanderman-piran2015} $R_{\rm GRB}$($\gtrsim 10^{52}$ ergs)$\sim0.1$ Gpc$^{-3}$\,yr$^{-1}$, composing about 1\% of all SGRBs {\it that point towards Earth}. 
%At first glance it seems that, if the luminosity function derived by ref\cite{wanderman-piran2015} (and others) is correct, GW170817 produced an unusually luminous SGRB, and
This suggests either that we were extremely lucky in observing such an event or  that all such luminous events are more narrowly beamed than events of smaller $L_{iso}$, and do not typically point towards Earth. For example, if GW170817, with an opening angle of $\sim0.05$ rad, is representative of events of $L_{iso} \sim10^{52}$ erg s$^{-1}$, it would imply that there are 1000 events with such luminosity that point away, for every SGRB-producing event that points toward Earth, i.e. a rate of $\sim$100 Gpc$^{-3}$\,yr$^{-1}$ for GW170817-like events. This rate is about 3\%--30\% of all the neutron star binary merger rate\cite{abbott2017a}, $R_{\rm BNS}=1540_{-1220}^{+3200}$ Gpc$^{-3}$\,yr$^{-1}$, and would imply that the true fraction of high luminosity SGRBs is much higher than observed at Earth.  An anticorrelation between the jet opening angle and its isotropic equivalent energy is one possible cause for such a relationship, and rather naturally follows if the total energy of different events varies less than their beaming. This can be easily tested with a small number of future events with off-axis afterglow emission.

\clearpage
\section*{References}
\bibliography{refs}

%%%%%%%%%%%%%%%%%%%%%%%%%%%%%%%%%%%%%%%%%%%%%%%%%%%%%%%%%%%%%%%
% MAIN FIGURES
%%%%%%%%%%%%%%%%%%%%%%%%%%%%%%%%%%%%%%%%%%%%%%%%%%%%%%%%%%%%%%%
\clearpage

%\ref{fig:light_curves}
\clearpage
\includegraphics[width=6.5in]{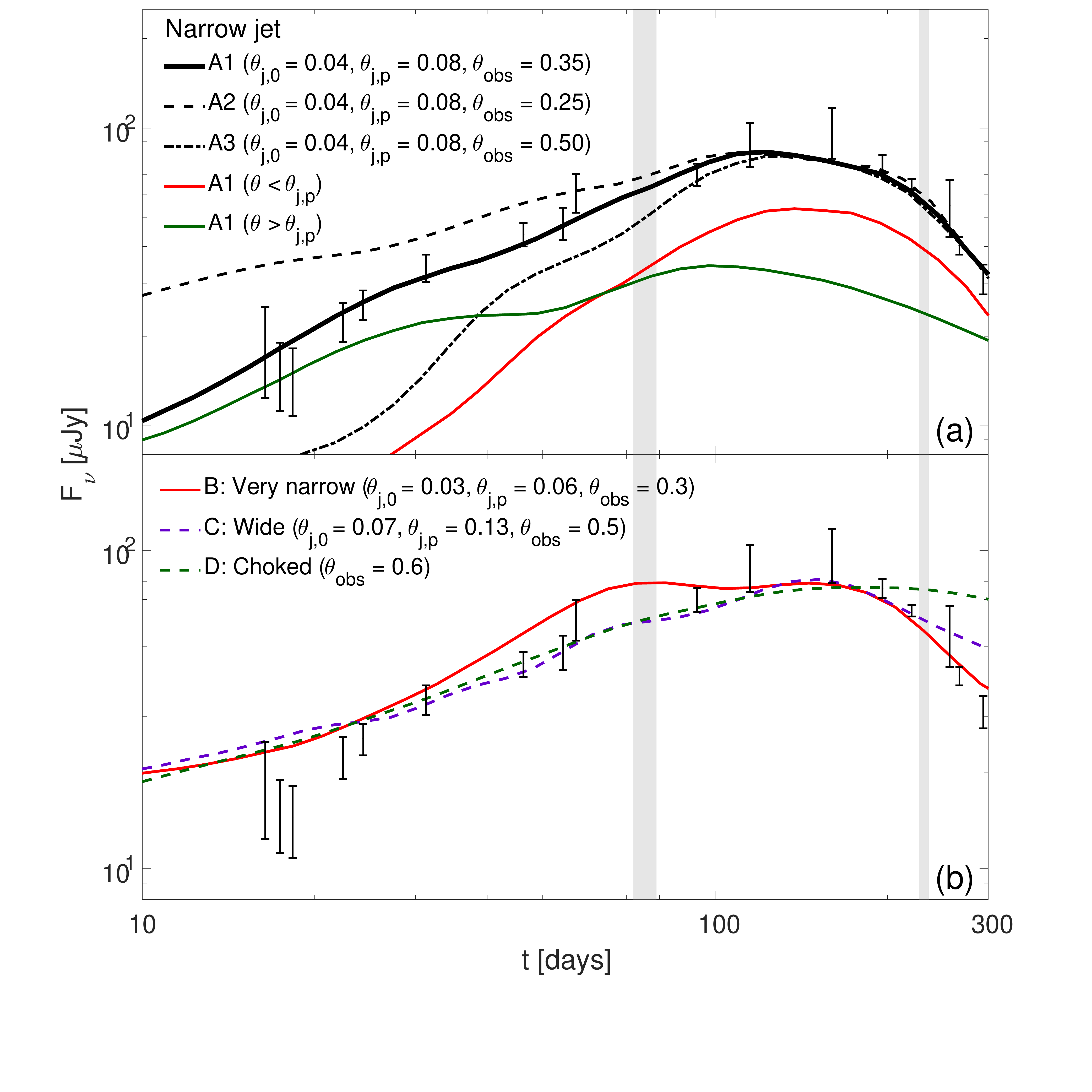}

\clearpage
\begin{figure}
\caption{{\bf Radio 3 GHz light curves of several representative simulated models.} The black errorbars (1$\sigma$) are the 3 GHz flux density values of GW170817. The grey shaded regions denote the VLBI epochs: 75d and 230d post-merger. {\it Panel (a)}: A narrow jet with an initial opening angle $\theta_{j,0}=0.04$ rad (2.3\deg), total energy $E=10^{50}$ erg, and isotropic equivalent energy $E_{iso}=10^{53}$ erg at the core, as observed at three different viewing angles (models $ \A1 $ --$ \A3 $). For all light curves, we take $\epsilon_e=0.1$ and $p=2.16$ and vary the energy fraction of the magnetic field $\epsilon_B$, and the external density (assumed to be constant in space), $n$, to obtain a best fit to the light curve. The opening angle of the jet core at the time of the peak is $\theta_{j,p}=0.08$ rad. The model that we find to fit best both the light curve and the images is at a viewing angle $\theta_{obs}=0.35$ rad ($\epsilon_B=10^{-4}$, $n=6 \times 10^{-4} {\rm~cm^{-3}}$). The red line shows the contribution of emission from the jet core ($\theta < \theta_{j,p}$) and the green shows the cocoon emission. The fit to the observations is obtained only in a rather narrow range of viewing angles. For smaller angles (e.g., $\theta_{obs}=0.25$ rad, $\epsilon_B=2 \times 10^{-4}$, $n=10^{-4} {\rm~cm^{-3}}$), the light curve rises too slowly and the image centroid moves too far, while at larger angles (e.g, $\theta_{obs}=0.5$ rad, $\epsilon_B=8 \times 10^{-5}$, $n=6 \times 10^{-3} {\rm~cm^{-3}}$), the light curve rises too quickly and the image centroid motion is too small.  {\it Panel (b)}: Light curves of three other models. Model $ \B $: Another narrow jet with a lower energy, $\theta_{j,p}=0.06$ rad, $E=10^{49}$ erg, $E_{iso}=2 \times 10^{52}$ erg ($\epsilon_B=4 \times 10^{-5}$, $n=7 \times 10^{-3} {\rm~cm^{-3}}$) at $\theta_{obs}=0.3$ rad, which provides a reasonable fit to the data.  Model $ \C $: A wider jet with $\theta_{j,p}=0.13$ rad. Even for $\theta_{obs}=0.5$ rad the light curve does not decay fast enough to be consistent with the most recent data points. At this viewing angle also the images centroid moves too slow. Model $ \D $: A model of a choked jet. The light curve does not decay fast enough after the peak and the image motion, while being superluminal, is very slow compared to the observations. Note that in all the models that we considered, the spectrum between the radio and the X-ray is a constant power-law (cooling and self-absorption do not affect this spectral range) and therefore models that fit the radio 3 GHz data, fit the entire afterglow observations from radio to X-ray. See methods for details.}
\label{fig:light_curves}
\end{figure}

\clearpage
\includegraphics[width=6in,trim={5cm 2cm 7cm 1.7cm},clip]{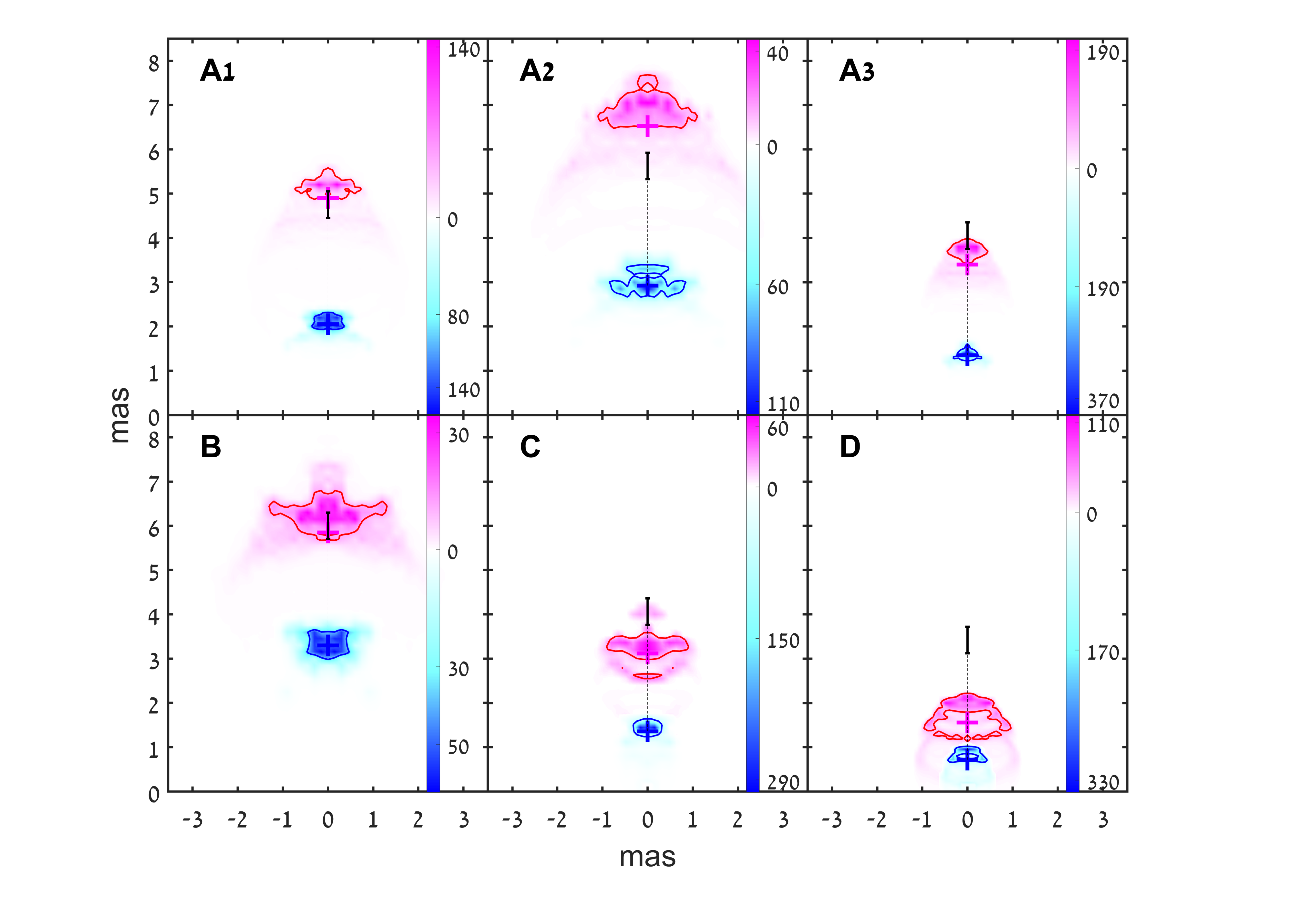}

\begin{figure}
\clearpage
\caption{{\bf Synthetic radio images.} Each panel shows two colormaps of the flux density ($\mu$Jy mas$^{-2}$), one at 75d (blue color palette) and one at 230d (magenta color palette) for the models $ \A1 $--$\A3, B, C $ and $\D$ shown in Figure~\ref{fig:light_curves}. The position at the time of merger is $x=y=0$, while the blue and magenta crosses mark the flux centroid at 75d and 230d respectively. The 50\% flux containment contours are also shown at the two epochs. The black dashed line marks the direction of centroid motion and the black solid segment denotes the motion consistent with the VLBI observations within 1$\sigma$, $2.7 \pm 0.3$ mas. Only models A1 and B, which are of narrow jets ($\theta_{j,p}<0.1$ rad) observed at angles of 0.35 rad and 0.3 rad, show centroids motions that are consistent with the observations (2.8 and 2.6 mas, respectively). These are also the models that provide the best fits to the light curve.  The centroid motion between the two epochs of successful jet models with larger opening angle, A3 and C (0.5 rad), is too small (2.1 and 1.7 mas respectively) while that of model A2 (0.25 rad) is too large (3.5 mas). The choked jet model, D, is much too small (0.7 mas). In all the successful jet models larger viewing angles lead to more compact images. The observed images were unresolved with an upper limit  on the width parallel to the centroid motion of about 1 mas ($1\sigma$). Models A1, A3 and C ($\theta_{obs} \geq 0.35$ rad) are consistent with this limit, model B ($\theta_{obs} = 0.3$ rad) is marginal, and model A2 ($\theta_{obs} = 0.25$ rad) is too extended. See Figure~\ref{fig:light_curves} and Methods for further details of the various models and their fitting to the VLBI data.}
\label{fig:images}
\end{figure}

\clearpage
\includegraphics[width=6in]{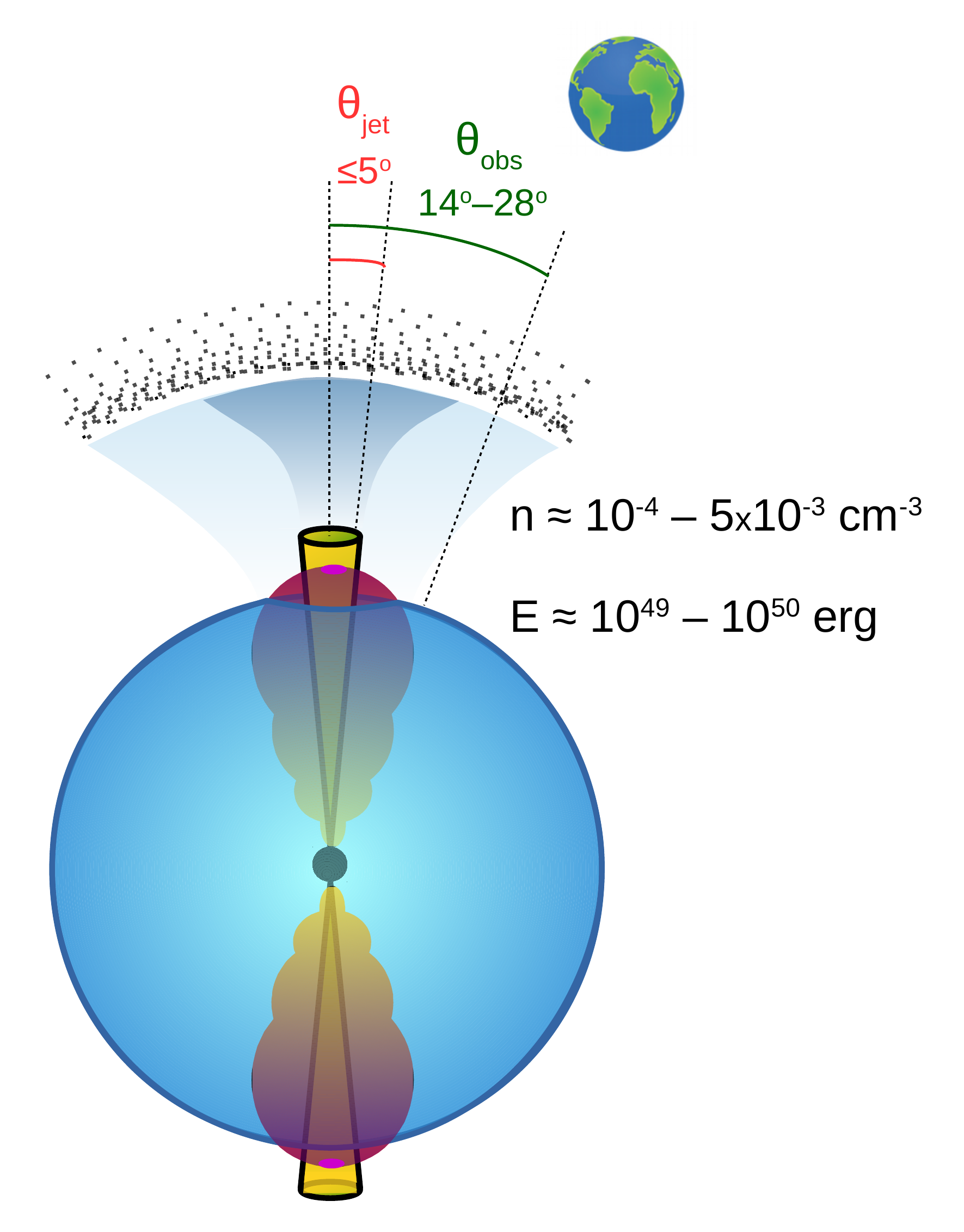}

\clearpage
\begin{figure}
\caption{{\bf Schematic illustration showing the physical and geometrical parameters derived for GW170817.} 
GW170817 has a successful jet (yellow) that drives a cocoon (red) through interaction with the dynamical ejecta (blue). This scenario is the same as scenario "E" in our previous work\cite{mooley2018} and consistent with structured jet models. The shock-breakout from the cocoon likely produced the gamma-ray signal and the cocoon's interaction with the ISM produced the early-time (up to $\sim$2 months post-merger) radio and X-ray emission. The relativistic core of the jet has a half-opening angle ($\theta_{\rm jet}$) of $\leqslant$5\deg. 
%The cocoon has a half opening angle between 15o-20o. 
The Earth is located 14\deg--28\deg away (the viewing angle, $\theta_{\rm obs}$) from the core of the jet. GW170817 most likely gave rise to a short gamma-ray burst pointing at such an angle away from the Earth. The interaction between the jet and the ISM produced the late-time radio and X-ray emission. Our VLBI measurement suggests that the Lorentz factor of the jet at 150 days post-merger (i.e. at the peak of the radio light curve, when the core of the jet came into view) is $\Gamma \approx 4$. The total energy (E) of the jet and cocoon system is between 10$^{49}$--10$^{50}$ erg. The density (n) of the circum-merger environment is between $10^{-4}-5\times10^{-3}$ cm$^{-3}$.}
\label{fig:schematic}
\end{figure}

% \clearpage
% \begin{figure}\label{fig:polarization}
% \centering
% \includegraphics[width=6.5in]{fig5.pdf}
% \caption{\bf{The linear polarization of the best-fitted model for different b values.}}
% \end{figure}

\clearpage
\begin{methods}

\section{Observations, Data processing \& Basic analysis}
In order to establish the size and morphology of the faint radio afterglow of GW170817, we obtained Director's Discretionary Time (program ID BM469) to observe with the High Sensitivity Array (HSA). The HSA antennas included the ten Very Long Baseline Array (VLBA) dishes, the phased Karl G. Jansky Very Large Array (VLA), and the Green Bank Telescope (GBT), although not all stations were present in all observations.  The maximum baseline was typically 7,500--8,000 km.

\subsection{VLBI Observations}
We observed GW170817 with the HSA over four epochs between 2017 September -- 2018 April. Each epoch consisted of 2--4 observations carried out over a period of up to 10 days, with approximately three hours of on-source time on GW170817 per day. The choice of the observing frequency was informed by the results from the VLA monitoring of the radio light curve, the desired angular resolution, and the ease of scheduling on the telescopes.  In all epochs, a total bandwidth of 256 MHz was sampled in dual polarisation at 2-bit precision. Depending on the observing frequency, the recorded bandwidth was broken into eight 32 MHz wide bands, or two 128 MHz wide bands. A summary of the observations is given in Table~\ref{tab:obs}.

The first epoch was undertaken at L band (central frequency 1550 MHz) 37 -- 38 d post-merger. No fringes were seen on the GBT on one of the two observing days due to an unknown technical issue, considerably reducing  overall sensitivity at this epoch.
The second epoch was carried out in S band (central frequency 3200 MHz), 51 -- 52 d post-merger. However, a misconfiguration of the VLA correlator on both days meant that phased VLA data was practically unusable, and hence sensitivity was severely impacted.  The third epoch was observed at C band (central frequency 4540 MHz) 72 -- 79 d post-merger.  The fourth epoch was likewise observed at C band 227--236 d post-merger, utilising only the VLBA and VLA as the GBT was unavailable.

Each observation was structured around an 8 minute cycle as follows.  We used the source J1258-2219 (a $\sim$1 Jy flat-spectrum source, separated by 2.8 degrees from GW170817) as the primary delay and gain calibrator, visiting it twice per cycle during first three epochs, and once per cycle in the fourth epoch observations. J1312-2350, a 20 mJy source separated by 0.8 degrees from GW170817, was used as a secondary phase calibrator, and was visited once per cycle in the first three epochs, and twice per cycle in the fourth epoch observations.  J1258-2219 was additionally used to determine phase solutions for the VLA once per cycle.  A single scan on 3C286 was included at the end of each observation to allow flux calibration of the commensally-recorded VLA interferometer data.  For the C band (4.5 GHz) epochs only, we included three scans on the blazar OQ208 (B1404+286) over the course of each observation to enable polarization calibration to be determined and applied.

\subsection{VLBI Data Processing}
We followed standard data reduction procedures for HSA data using the AIPS software package \citemethod{greisen2003}. For all calibration steps that involve a sky source (fringe-fitting, leakage, and self-calibration) we used a model of the source that was iteratively refined over several passes of the entire data reduction pipeline.

The data was loaded using “FITLD” and a priori amplitude corrections were applied using “ANTAB” and “ACCOR”. We note that an issue with the VLA automatic gain control was uncovered whereby the phased VLA data exhibited large short-term amplitude variations; this could be (and was) largely mitigated by using a per-integration solution for the auto-correlation based corrections with “ACCOR”, but small residual variations which were weakly detrimental to sensitivity remained.  This problem was fixed prior to the fourth observational epoch.  “CLCOR” was used to correct for parallactic angle rotation and to apply the most accurate available values for Earth Orientation Parameters. TECOR was used to correct for ionospheric propagation effects, using the “igsg” model available from {\tt ftp://cddis.gsfc.nasa.gov/gps/products/ionex}.
We then calibrated the time-independent delays and the antenna bandpass using “FRING” and “BPASS”; in the first two epochs using a scan on the primary calibrator J1258-2219, while in the third and fourth epochs we used OQ208.

For the third epoch at 4.5 GHz only, we calibrated the cross-polar delays and instrumental polarization leakage using the  tasks “FRING” and “LPCAL” and the source OQ208.  This step was essential due to the large ($\sim$30\%) leakage at the GBT at this frequency. “LPCAL” solves for a single leakage value per subband, while the GBT polarisation leakage varies across the 128 MHz subband; accordingly, we split each 128 MHz subband into 4$\times$32 MHz subbands to allow a coarse frequency dependence to the leakage solutions.

We solved for time dependent delays using “FRING” on the primary gain calibrator J1258-2219, followed by self-calibration on this source using “CALIB”, obtaining a single solution per subband, per scan.  Finally, we improved the phase calibration using self-calibration on the secondary gain calibrator J1312-2350, deriving a single frequency-independent solution per scan.

At each stage, the solutions from the SN table were applied to the CL table using “CLCAL”. The final CL table was applied to the target using “SPLIT”. The target was then exported in UVFITS format using “FITTP” and imaged using “difmap”\citemethod{shepherd1997}. 

\subsection{VLA/VLBI Interferometric data processing}
We processed using VLA cross-correlated data (with the WIDAR correlator) using a custom-developed pipeline, which incorporates manual flagging, and standard interferometric data calibration techniques in CASA. The imaging was done with the CASA task clean with natural weighting, choosing an image size of 4096 pix $\times$ 4096 pix and cell size of 0.5 arcsec.

The VLA-only data gives the GW170817 flux densities of $56\pm8$ \ujybeam, $54\pm8$ \ujybeam and $45\pm7$ \ujybeam for the three observations of the third epoch at 4.5 GHz. All three observations combined give $55\pm5$ \ujybeam. For the four observations of the fourth epoch, the flux density values are $55\pm8$ \ujybeam, $46\pm8$ \ujybeam, $48\pm6$ \ujybeam and $46\pm6$ \ujybeam, while all four observations combined give $48\pm4$ \ujybeam.

\subsection{Flux comparison between the VLBI and VLA interferometric data}
A comparison between the flux densities measured in the VLA-only interferometric data and those measured in the VLBI data (see Extended Data Table 1) implies that, within 1$\sigma$ uncertainties (typically 10\% of the source flux density), no flux is being resolved out in the VLBI data.

\subsection{Model fits and parameter estimations}
Difmap\cite{shepherd1997} was first used to produce a "dirty" (un-deconvolved) image from the concatenated data from each epoch, as well as the individual observations within each epoch. In the first two epochs, there was substantial loss of sensitivity due to technical issues and the source was not detected.  We place 5 upper limits of 40 \ujybeam (1.6 GHz, day 38) and 60 \ujybeam (3.2 GHz, day 52), respectively on the flux densities of GW170817, and do not consider these epochs further.

In the third and fourth epochs, a radio counterpart to GW170817 can clearly be seen in the dirty images for the concatenated datasets, and the source can also be seen (albeit at low S/N) in the individual observations.  Initially, we fit the data in the visibility plane using a single circularly symmetric gaussian model component.  Whilst likely an over-simplification of the true source structure, this has the advantage of being fast and simple to fit, while providing an accurate estimate of the flux centroid position.  After model fitting, we read the resultant clean image into AIPS and used the task JMFIT to fit an elliptical gaussian in the image plane.  Compared to model fitting, this has the advantage of providing well-constrained estimates of the uncertainty of the key parameters of interest\citemethod{condon1997}.  In the third epoch (75 days), the best-fit values of flux density and position are $58\pm5$ \ujybeam and RA=13:09:48.068638(9), Dec=-23:22:53.3909(4) respectively.  The uncertainties given here are purely statistical; we consider systematic contributions in the following sections.  The best-fit size was a full-width half-maximum (FWHM) of 0.0 mas; i.e., the source was modeled as a point source.  At day 230, the best-fit values of flux density and position were $48\pm6$ \ujybeam, RA=13:09:48.068831(11) Dec= -23:22:53.3907(4) respectively, and the best-fit deconvolved size was 0.7 mas, although an unresolved source could not be excluded.  The images of the source at 75 days and 230 days are shown in Extended Data Figure 1.

\subsection{Estimating systematic contributions to flux density and position uncertainties}
The absolute calibration of flux densities in VLBI maps is typically challenging due to the fact the sources compact enough to be visible at milliarcsecond resolution typically show evolution on a timescale on months to years.  In cases where only a priori amplitude calibration can be performed, the accuracy of the flux density scale of a VLBI image is typically assumed to be of order 20\%. In this case, we are able to use the contemporaneous VLA data to establish an absolute flux density scale, using the calibrator sources J1312-2350 and J1258-2219 (under the assumption that these sources do not have significant structure on scales larger than that resolvable by our VLBI observations).  After adjusting the VLBI amplitude scale to produce the closest match to these two sources, the residual differences are typically 10\% for each observation, and hence systematic uncertainties on our measured values of flux density for GW170817 are comparable to our statistical uncertainties.

Similarly, for our image centroid positions, we must consider the possibility of systematic position shifts between epochs due to calibration errors, in addition to the limiting precision attainable based on the image resolution and S/N. We neglect systematic errors due to the uncertainty in the calibrator reference position, since this would affect both epochs equally. Given the relatively close proximity of our calibrator source J1312-2350 to GW170817 (0.8 degrees), we expect any systematic errors that vary between epochs to be at most a small fraction of the synthesized beam size. Astrometric simulations\citemethod{pradel2006} suggest a typical systematic error for a single observation with the VLBA of 0.07 mas in right ascension and 0.25 mas in declination for our observing conditions (declination $-26$ degrees, angular separation 0.8 degrees).  However, these simulations do not include the effect of the ionosphere, which could treble the systematic error at an observing frequency of 4.5 GHz under typical conditions.  Countering this somewhat, our epochs consist of 3--4 observations spread over $\sim$7 days, and systematic errors (in particular those due to the ionosphere) are likely to be only weakly correlated over this timescale.  Based on these considerations, we estimated the systematic position uncertainty to be 0.15 mas in R.A. and 0.5 mas in declination, and added this value in quadrature with the formal position fit errors at each epoch. 

In order to verify this expectation, we repeated the data reduction for the third and fourth epochs after shifting the phase center of our target field to the position of the NGC 4993 low-luminosity AGN.  This source is separated by 10.3 arcseconds from GW170817, and hence falls outside the field of view of the phased VLA; accordingly, the VLA was flagged before imaging.  The positions obtained for the AGN have a separation of 0.05 mas in right ascension and 0.5 mas in declination (see Extended Data Figure 2).  This is consistent with both their statistical uncertainties and our estimate for the systematic errors derived above. The AGN flux density is consistent with a constant value ($0.25 \pm 0.02$ mJy and $0.29 \pm 0.03$ mJy in the third and fourth epochs respectively, where the 1$\sigma$ uncertainties are purely statistical).

\section{Comparison between the VLBI data and synthetic images}

In order to compare the generated models with our VLBI data, we converted the simulated images (example images shown in Figure~\ref{fig:images}; for details of the simulations see the next section) into difmap models consisting of point sources at the center of each non-zero pixel in the simulated image, and performed model fitting in the visibility plane.  The rotation, translation, and total flux density of the image were taken as free parameters, although we used the approximate positions and flux densities from our earlier fitting of circular gaussian components to restrict the ranges of parameter values over which we searched.  For each model, we recorded the $\chi^2$ obtained at the best-fit values for rotation, translation, and total flux density.

Because the signal-to-noise of each individual visibility measurement is very low, determining the increase in $\chi^2$ that indicates a significant discrepancy between models is not straightforward.  Previous authors have often relied on visual inspection of images and visibility data in order to determine model goodness-of-fit\citemethod{tzioumis1989,tingay2001}.  Due to the low signal-to-noise ratio of our target image, we have taken a different approach.  First, we used an image plane fit to determine the position errors in the image plane using the dataset fit with a circular gaussian component, which is a well-understood process\cite{condon1997}.  Second, we perturbed the position of the circular gaussian model component by up to $\pm$3$\sigma$ in right ascension and $\pm$3$\sigma$ in declination, and recorded the change in $\chi^2$ at offsets of 1, 2, and 3$\sigma$.  A consistent increase in $\chi^2$ was seen regardless of the direction of the positional perturbation.  
%Finally, we fitted other models based on the hydrodynamic simulations to the data: any model producing a $\chi^2$ within plus or minus this perturbation value of the $\chi^2$ of the best circular gaussian fit was considered equivalently good to the circular gaussian, above the upper limit was considered significantly worse, and below the lower limit was significantly better.
Finally, we fitted other models based on the hydrodynamic simulations to the data and recorded the $\chi^2$ in each case.  The reference positions for a given model were allowed to vary between the day 75 and day 230 datasets by up to the amount of our estimated systematic position uncertainty of 0.15 mas in R.A. and 0.5 mas in Declination.  By comparison to the set of $\chi^2$ values obtained from the perturbed circular gaussian fits, we estimated the consistency of each hydrodynamic model with the best-fit circular gaussian model.

%{\bf ATD ADDED (note that I also changed a couple of +/- to $\pm$ in the preceding paragraph of this section, and changed the text saying "chi-squared" to $\chi^2$ throughout.  Suggested added text about the size limits now follows in bold, and I also tweaked the final sentence of this section, it is also bolded.

In addition to fitting the actual synthetic images, we first produced an estimate of the maximum source extent, by finding the largest circular and elliptical gaussian sources that produced a $\chi^2$ that did not deviate by more than 1\,$\sigma$ from the best circular gaussian fits.  For the epoch at day 75 and day 230, the largest circular gaussian source was 1.1 and 1.2 mas in diameter respectively.  The best-fit elliptical gaussian converged to an unphysical one-dimensional source for each epoch, with an upper limit on the major axis of 12 mas and 9 mas for day 75 and day 230 respectively. In both cases the best-fit position angle was approximately aligned with the beam major axis and hence approximately perpendicular to direction of source motion.  Tighter limits on the maximum size can be obtained if the axial ratio of the elliptical gaussian source is constrained to a physical value: for instance, in the case of the day 230 dataset, the largest source permitted with an axial ratio of 4:1 has size 3.9 mas $\times$ 0.9 mas. Hence, the source size parallel to the direction of motion is relatively well constrained.

None of the synthetic images produced a $\chi^2$ significantly better than a simple circular gaussian in either epoch (unsurprising, given that the source was consistent with being unresolved in both cases).  %Generally, we found that most models capable of producing the $\sim$2.7 mas of positional offset between days 75 and 230 predicted a source that was too extended by day 230, while models which remained sufficiently compact at day 230 did not show a sufficiently large positional offset.  The best-fitting model (narrow jet viewed at 0.35 radians) was able to produce the expected positional shift between epochs: with a constant reference translation and rotation, it produced an acceptable fit to the  day 75 epoch (equivalent $\chi^2$ to the circular gaussian), and a marginally acceptable fit to the day 230 epoch ($\chi^2$ increased to that obtained when shifting the circular gaussian position 2.5$\sigma$ away from the best-fit position). 
Generally, we found that as the positional offset between days 75 and 230 increased, the "best-fit" source size at day 230 also increased and was often inconsistent with the observed source compactness.  This disfavoured models at low viewing angles.  Conversely, models at large viewing angles were incapable of producing a sufficiently large positional offset.

The best-fitting model (narrow jet viewed at 0.35 radians, model A1 in Figures~\ref{fig:light_curves} and \ref{fig:images}) was able to produce the expected positional shift between epochs: with a constant reference translation and rotation, it produced an acceptable fit to both the day 75 epoch ($\chi^2$ increase equivalent to a 0.9$\sigma$ position offset for the circular gaussian) and the day 230 epoch ($\chi^2$ increase equivalent to a 1.3$\sigma$ position offset for the circular gaussian). Among the other models, only one (model B, the very narrow jet viewed at 0.3 radians) remained consistent within 2$\sigma$ for both epochs.  For all other models, the discrepancy with the best-fit circular gaussian exceeded 2$\sigma$ in one or both epochs.  As can be seen in Figure~\ref{fig:light_curves}, models A1 and B are also those that best fit the light curve.
%{\bf ATD: need to update numbers if Figures 2 and 3 are updated.}}  
%\refg{Any small changes to the jet characteristics that concentrated a greater proportion of the emission at day 230 into a smaller region while maintaining the positional separation between day 75 and day 230 would improve the consistency of the fit.}

\section{Numerical hydrodynamic simulations}

To characterize the properties of different models we carry out relativistic hydrodynamical simulations of various setups, followed by a post processing numerical calculation\cite{nakar2018} of their afterglow light curve and observed images at 75 and 230 days.
In particular we run different type of models to see which have the potential to fit the entire data set of both the light curve and the image characteristics, i.e. the flux centroid movement and the image size constraints.

Our setup includes three components: the jet, a core of cold massive ejecta and a fast ejecta tail.
Each component of the ejecta expands homologously and has a density profile of
\begin{equation}
\rho(r,\theta) = \rho_0 r^{-\alpha} \Big(\frac{1}{4}+{\rm sin}^\beta\theta\Big)~,
\end{equation}
where the normalization $ \rho_0 $ is determined by the total ejecta mass and $ \alpha $ and $ \beta $ which differ between models, dictate the radial and angular structures, respectively. 
However, our main focus was on scanning the jet's properties such as luminosities, opening angles, injection and delay times. While some of the jets successfully break out from the ejecta if their properties allow, others may be choked inside it.
We ran about ten different models, here we present four representative models that demonstrate how the different characteristics of the jet affect the observed outcome.  The first two models are narrow jets which are found to fit all the observed characteristics-the gradual rise of the flux, the short plateau at the peak followed by a fast decline and the large flux centroid motion between the two image epochs. In addition we also present a wider successful jet and a choked jet.
The full setup is given in Extended Data Table~\ref{tab:models_parameters}.

A full description of the  hydrodynamic simulations is given in our previous work\cite{nakar2018}. Briefly, for each model we use three different simulations. The first one which includes the jet propagation inside the core ejecta is performed in 3D to avoid the numerical plug artifact\citemethod{gottlieb2018a}. 
The second simulation includes the outflow evolution inside the tail ejecta and after breaking out of it until reaching the homologous phase. This simulation is modeled in 2D as previously we showed\citemethod{gottlieb2018b} that after breakout the plug artifact is no longer a concern, and 2D and 3D simulations become similar. Finally, the third simulation begins when the afterglow becomes important and ends after it decays.

For the relativistic hydrodynamical simulation we use the public code PLUTO\citemethod{mignone2007} v4.0 with an HLL Riemann solver and we apply an equation of state with adiabatic index of 4/3.
The setup of models $ \A $ and $ \B $ is as follows.
The grid setup of the first 3D Cartesian simulation has three patches in x and y axes and two patches on the z axis. On x and y the inner patch spans from $ -2\times 10^8\cm $ to $ 2\times 10^8\cm $ with 30 uniform cells. The outer patch is from $ |2\times 10^8\cm| $ to $ |3\times 10^{10}\cm| $ with 400 cells that are distributed logarithmically. On the z-axis the first patch is uniform from $ 4.5\times 10^8\cm $ to $ 10^{10}\cm $ with 200 cells followed by a logarithmic patch of 400 cells until $ 4\times 10^{10}\cm $. We convert the 3D output of the first simulation to an axisymmetric grid\citemethod{gottlieb2018b}, which is the initial setup of the second simulation for which the setup is as follows. The first two patches on r and z axes correspond to the 3D setup. We add another patch on each axis from $ 3\times 10^{10}\cm $ ($ 4\times 10^{10}\cm $) on the r (z) axis, to $ 6\times 10^{11}\cm $ with 1200 logarithmic cells.

For the third simulation which includes two patches on each axis, we use the output of the second simulation. The first patch corresponds to the second simulation grid with 800 uniform cells until $ 6\times 10^{11}\times R\cm $ on each axis. The second patch on each axis stretches to $ 10^{14}\times R\cm $ with 6000 logarithmic cells. As the simulation is dimensionless, we use $ R $ as a scaling length factor\cite{nakar2018}, $ R $ also determines the ISM density which is set to be $ \rho_{ISM} = 5\times 10^{-12}\rm{gr}(R\times \cm)^{-3} $ in simulation $ \A $ and $ \rho_{ISM} = 8\times 10^{-12}\rm{gr}(R\times \cm)^{-3} $ in simulation $ \B $. Each viewing angle fit requires a different $ R $. The best fits for $ \theta_{obs} = 0.25, 0.35, 0.45 $ in simulation $ \A $ are obtained at $ R =  3\times 10^5, 1.7\times 10^5, 8.3\times 10^4 $, respectively, and for $ \theta_{obs} = 0.3 $ in simulation $ \B $ it is $ R = 5\times 10^5 $.

The setup of simulations $ \C $ and $ \D $ was described previously\cite{nakar2018} (simulation $ \D $ is identical to the successful jet scenario, except for the engine time), and the only difference here is that for the outer patch in the third part we use a high resolution of 4000 cells rather than 2500 cells originally. The scaling of the third part of the simulation is determined by $ n = 4\times 10^{-2}~\rm{cm}^{-3} $ and $ n = 4.5\times 10^{-3}~\rm{cm}^{-3} $ in $ \C $ and $\D $ respectively.

Finally, we verify that each of the three simulation meets the required resolution to reach convergence.
We first compare the resolution of the first two simulations, from the jet launch until reaching the homologous phase, with previously-published simulations\citemethod{gottlieb2018a} for which convergence tests have been taken.
The resolution of the 3D simulation which handles the jet propagation inside the ejecta is comparable with that of the inner parts of theirs. The sequential 2D simulation has naturally a higher resolution compared with the outer parts of the 3D grid presented previously\citemethod{gottlieb2018a}.
For convergence of the third part in which the outflow interacts with the ISM, we perform another set of simulations with 2/3 the resolution aforementioned. We find that both the light curves and the images for the relevant viewing angles remain essentially unchanged with the increase in resolution.

\section{Details of the simulation that provides the best fit to the data}
Our simulation that provides the best fit to the data is of a jet with a 0.08 rad (4\deg) opening angle, at the time of light curve peak, that is observed at a viewing angle of $\theta_{\rm obs}=0.35$ rad (20\deg). 
In this simulation a relativistic jet is injected into the sub-relativistic merger ejecta. The jet is followed during its propagation through the ejecta, the formation of the cocoon and the breakout of the jet and the cocoon from the dynamical (sub-relativistic) ejecta. The simulation then continues to follow the interaction of the outflow (jet+cocoon) with the ISM.  When this interaction starts, the jet opening angle is 0.04 rad. The cocoon dominates the observed radio emission during the first $\sim$60 days, and after this time the jet dominates. The jet expands sideways slowly during its interaction with the ISM, reaching an opening angle of 0.08 rad after $\sim$150 days at the light curve peak.
On day 75, the Lorentz factor of the observed region is $\Gamma\approx4$, which steadily drops to $\Gamma\approx3$ by day 230. 

\section{Constraining the jet energy and the external density}
The gamma-ray signal from GW170817 had an isotropic equivalent energy of $5\times 10^{47}$ erg.
The afterglow suggests that this energy is not representative of the jet energy.
This is consistent with models for the gamma-ray emission\citemethod{lazzati2017, eichler2017,kathirgamaraju2018,kasliwal2017,gottlieb2018b,bromberg2018,pozanenko2018}.
Therefore, in order to constrain the jet energy and external density, we use the constraints on the geometry of the outflow together with the observed afterglow light curve to constrain the outflow energy.
We use the standard afterglow model, where a narrow ultra-relativistic jet drives a blast wave into the external medium which radiates in synchrotron emission to produce the radio and X-ray afterglow. Before interacting with the external medium the jet has an initial Lorentz factor $\Gamma_0$. This is also the initial Lorentz factor of the blast wave that it drives, which is constant at first until the blast wave accumulates 
 enough mass and starts decelerating. Its initial opening angle, $\theta_{j,0}$, is also constant until the Lorentz factor drops to
 $\sim 1/\theta_{j,0}$. At this point, if $\theta_{j,0} < 0.05$ rad it starts spreading sideways rapidly until $\theta_{j,0} \sim 
 0.05$  rad, at which point it starts spreading sideways more slowly\citemethod{granotpiran2012}. We have direct constraints only
 of $\Gamma$ and $\theta_j$ near the time of the peak of the light curve. We therefore can only put a lower limit on the initial
 Lorentz factor, $\Gamma_0>4$, and an upper limit on the initial opening angle $\theta_{j,0}<0.1$  rad. Moreover, given the fast
 spreading of the jet if $\theta_{j,0}<0.1$ rad and $\Gamma<1/\theta$, at the time that we observe the jet its opening angle is
 expected to be  $\theta_j \approx 0.05-0.1$ even if initially  $\theta_{j,0} \ll 0.1$ rad and its Lorentz factor is $\Gamma_0 \gg 4$.
 The Lorentz factor and the time of the peak provide a relation between the ambient medium density (assumed to be constant) and the
 jet isotropic equivalent energy\cite{nakar-piran2018}: $E_{iso} \sim 10^{52} \frac{n}{3\times 10^{-4} {\rm cm^{-3}}} $ erg. The flux is extremely
 sensitive to the Lorentz factor and we can use its value at the peak to constrain the density and the fraction of the internal
 energy that goes to the magnetic field\cite{nakar-piran2018}, $\epsilon_B$: $\frac{n}{3\times 10^{-4} {\rm cm^{-3}}} \left(
 \frac{\epsilon_B}{10^{-3}}\right)^{0.47} \sim \left(\frac{\Gamma}{3.5}\right)^{5.9}$, where we assume that 10\% of the internal energy goes to the accelerated electrons ($\epsilon_e=0.1$) and that their distribution power-law index is $p=2.16$. Allowing the least constrained parameter, $\epsilon_B$, to vary between $10^{-2}$ and $10^{-5}$  we find that the circum-merger density is $10^{-4} - 5 \times 10^{-3} {\rm cm^{-3}}$ and the jet isotropic equivalent energy is $E_{iso} \sim 3 \times 10^{51} - 10^{53}$ erg. Since the jet opening angle at this time is 0.05--0.1 rad and it contains a significant fraction of the total energy of the relativistic outflow (jet+cocoon), we find that the energy deposited by the merger in relativistic ejecta is $10^{49}-10^{50}$ erg. 
The confirmation of a successful jet in GW170817 also implies high isotropy of the magnetic field\citemethod{corsi2018}.

\end{methods}

%\section*{References}
\bibliographystylemethod{naturemag}
\bibliographymethod{refsmethod}

%%%%%%%%%%%%%%%%%%%%%%%%%%%%%%%%%%%%%%%%%%%%%%%%%%%%%%%%%%%%%%%
% ACKNOWLEDGEMENTS
%%%%%%%%%%%%%%%%%%%%%%%%%%%%%%%%%%%%%%%%%%%%%%%%%%%%%%%%%%%%%%%
\clearpage
\begin{addendum}
\item[Acknowledgements] The authors are grateful to the VLBA, VLA and GBT staff, especially Mark Claussen, Amy Mioduszewski, Toney Minter, Frank Ghigo, Walter Brisken, Karen O'Neill and Mark McKinnon, for their support with the HSA observations. The authors thank Vivek Dhawan and Paul Demorest for help with the observational issues with the VLBI system at the VLA. K.P.M would like to thank Amy Mioduszewski, Emmanuel Momjian, Eric Greisen, Tim Pearson and Shri Kulkarni for helpful discussions. The authors thank Mansi Kasliwal for providing critical comments on the manuscript. The authors also extend their thanks to the referees for providing useful comments. The National Radio Astronomy Observatory is a facility of the National Science Foundation operated under cooperative agreement by Associated Universities, Inc. K.P.M. is currently a Jansky Fellow of the National Radio Astronomy Observatory. K.P.M acknowledges support from the Oxford Centre for Astrophysical Surveys which is funded through the Hintze Family Charitable Foundation for some initial work presented here. EN acknowledges the support of an ERC starting grant (GRB/SN) and an ISF grant (1277/13). A.T.D. is the recipient  of  an Australian  Research  Council  Future  Fellowship (FT150100415). GH acknowledges the support of NSF award AST-1654815. AH acknowledges support by the I-Core Program of the Planning and Budgeting Committee and the Israel Science Foundation. AC acknowledges support from the National Science Foundation CAREER award \#1455090 titled "CAREER: Radio and gravitational-wave emission from the largest explosions since the Big Bang".
\end{addendum}

%%%%%%%%%%%%%%%%%%%%%%%%%%%%%%%%%%%%%%%%%%%%%%%%%%%%%%%%%%%%%%%
% AUTHOR INFORMATION
%%%%%%%%%%%%%%%%%%%%%%%%%%%%%%%%%%%%%%%%%%%%%%%%%%%%%%%%%%%%%%%
\clearpage

\begin{addendum}
\item[Author Contributions] KM, AD, SB, GH, DF coordinated the VLBI observations. AD, KM performed the VLBI data processing. OG, EN carried out the theoretical study including analytic calculations and numerical simulations, with some inputs from KH. KM, AD, EN, GH, DF wrote the paper. AC and AH compiled the references. AH, AD, KM compiled the methods section. OG, AD, AH, KM prepared the figures. All coauthors discussed the results and provided comments on the manuscript.

 \item[Competing Interests] The authors declare that they have no
competing financial interests.

 \item[Correspondence] Correspondence and requests for materials should be addressed to K.P.M. (email: kunal@astro.caltech.edu) and O. Gottlieb (email: oregottlieb@gmail.com).
 %, A.T.D. (adeller@astro.swin.edu) and O.G. (oregottlieb@gmail.com).
 
\item[Data Availability] All relevant (VLBI) data are available from the corresponding author on request. The VLA data (presented in Figure 2) are currently being readied for public release.

\item[Code Availability] The hydrodynamic simulations were done using the publically available code PLUTO.
 Radio data processing software: AIPS, DIFMAP, CASA.
\end{addendum}

%%%%%%%%%%%%%%%%%%%%%%%%%%%%%%%%%%%%%%%%%%%%%%%%%%%%%%%%%%%%%%%
% EXTENDED DATA
%%%%%%%%%%%%%%%%%%%%%%%%%%%%%%%%%%%%%%%%%%%%%%%%%%%%%%%%%%%%%%%

\newpage
% \section*{Extended Data}

\setcounter{figure}{0}
\renewcommand{\figurename}{Extended Data Figure}
\renewcommand{\tablename}{Extended Data Table}
\begin{table*}
\centering
\caption{{\bf Log of VLBI (HSA) observations}}
\label{tab:obs}
\begin{tabular}{llllllll}
\hline
\hline
Epoch & Date        & Time        & $\nu_c$  & BW    & $\Delta t$ & $F_\nu$    & Comments\\
      & (UT)        & (UT)        & (GHz)    & (MHz) & (days)     &($\mu$Jy/beam)  &  \\
\hline
1     & 2017 Sep 23 & 16.5h--22.5h & 1.6      & 256   & 37         & $<$40     & No fringes on the GBT\\
      & 2017 Sep 24 & 16.5h--22.5h &          &       & 38         & \\
\hline
2     & 2017 Oct 07 & 15.5h--21.5h & 3.2      & 128   & 51         & $<$60     & VLA mis-configured\\
      & 2017 Oct 08 & 15.5h--18.8h &          &       & 52         &           & VLA mis-configured\\
\hline
3     & 2017 Oct 28 & 14.5h--20.5h & 4.5      & 256   & 72         & $58\pm5$  & \\
      & 2017 Oct 29 & 14.5h--20.5h &          &       & 73         & \\
      & 2017 Nov 04 & 14.0h--20.0h &          &       & 79         & \\
\hline
4     & 2018 Apr 01 & 04.5h--10.5h & 4.5      & 256   & 227        & $48\pm6$  & VLBA+VLA\\
      & 2018 Apr 02 & 04.5h--10.5h &          &       & 228        &           & VLBA+VLA\\
      & 2018 Apr 04 & 04.5h--10.5h &          &       & 230        &           & VLBA+VLA\\
      & 2018 Apr 10 & 04.5h--10.5h &          &       & 236        &           & VLBA+VLA\\
\hline
\multicolumn{8}{p{6.5in}}{Table Notes: $\nu_c$ is the center observing frequency, BW is the effective bandwidth after RFI excision, $\Delta t$ is the time post-merger, and $F_\nu$ is the peak flux density of GW170817.}
\end{tabular}
\end{table*}

\clearpage

\begin{table}
	\setlength{\tabcolsep}{10pt}
	\centering
	\begin{tabular}{ | l | c  c | c  c | }
		\hline
		Model type & \multicolumn{2}{c|}{Narrow jets} & Wider jet & Choked jet \\ \hline
		$  $Model & $ \A $ & $ \B $ & $ \C $ & $ \D $ \\ \hline
		$ L_j~(10^{50}\erg) $ & 1.4 & 0.6 & \multicolumn{2}{c|}{6.7} \\ 
		$ \theta_{inj}~ $ & $ 0.07 $ & $ 0.04 $ & \multicolumn{2}{c|}{$ 0.18 $} \\
		$ t_{inj}~(\s) $ & \multicolumn{1}{c}{0.2} & 0.3 & \multicolumn{2}{c|}{0.72}  \\
		$ t_{eng}~(\s) $ & \multicolumn{1}{c}{0.8} & 0.6 & 1.0 & 0.4 \\
		$ h_j $ & 200 & 400 & \multicolumn{2}{c|}{80} \\
		\hline
		$ M_{c}~(0.01\msun) $ & \multicolumn{2}{c|}{4} & \multicolumn{2}{c|}{5} \\
		$ M_{t}~(10^{-3}\msun) $ & \multicolumn{2}{c|}{1.6} & \multicolumn{2}{c|}{2.0} \\
		$ \alpha_{c} $ & \multicolumn{2}{c|}{2} & \multicolumn{2}{c|}{3.5} \\
		$ \alpha_{t} $ & \multicolumn{2}{c|}{14} & \multicolumn{2}{c|}{10} \\
		$ \beta $ & \multicolumn{2}{c|}{8} & \multicolumn{2}{c|}{3} \\
		$ v_{max,c}/c $ & \multicolumn{2}{c|}{0.2} & \multicolumn{2}{c|}{0.2} \\
		$ v_{max,t}/c $ & \multicolumn{2}{c|}{0.6} & \multicolumn{2}{c|}{0.8} \\
		\hline
	\end{tabular}
	\caption[Models parameters]{
		\rm{{\bf The initial setups of the simulation configurations $ \A-\D $.} The parameters of the jet are the total luminosity $ L_j $, opening angle upon injection $ \theta_{inj} $, injection delay time since the merger $ t_{inj} $, working engine time $ t_{eng} $ and specific enthalpy $ h_j $. The ejecta parameters are its mass $ M $, density radial power-law $ -\alpha $, density angular distribution $ \beta $ and front velocity $ v_{max} $. Each is given for the core with subscript $ c $ and tail with subscript $ t $.
			}
	}
	\label{tab:models_parameters}
\end{table}

\clearpage

\begin{figure}
\centering
\includegraphics[width=6.6in]{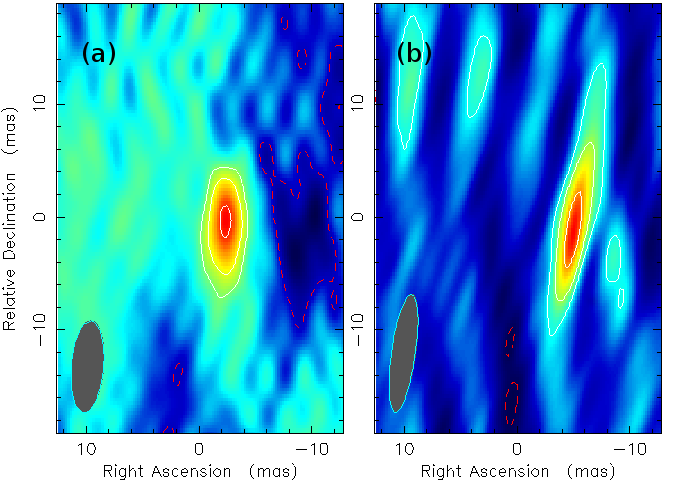}
\caption{{\bf VLBI images.}
The cleaned images (natural weighting; 0.2 mas pixel$^{-1}$) from the two epochs of VLBI, 75 d (panel a) and 230 d (panel b) post-merger. The center coordinates for these images are RA 13:09:48.069, Dec -23:22:53.39. The white contours are at 11, 22, and 44 \ujybeam in both images (red contour is $-11$ \ujybeam). The peak flux density of the sources is $58\pm5$ \ujybeam and $48\pm6$ \ujybeam in the two epochs respectively (image RMS noise quoted as the 1$\sigma$ uncertainty). The ellipse on the lower left corner of each panel shows the synthesized beam: [12.4, 2.2, -7] and [9.1, 3.2, -4] for the two epochs [major axis in mas, minor axis in mas, position angle in degrees].}
\label{fig:vlbi-images}
\end{figure}

\clearpage
\begin{figure}
\centering
\begin{tabular}{cc}
\includegraphics[trim=4.5cm 0cm 4.5cm 0cm, width=3in, clip]{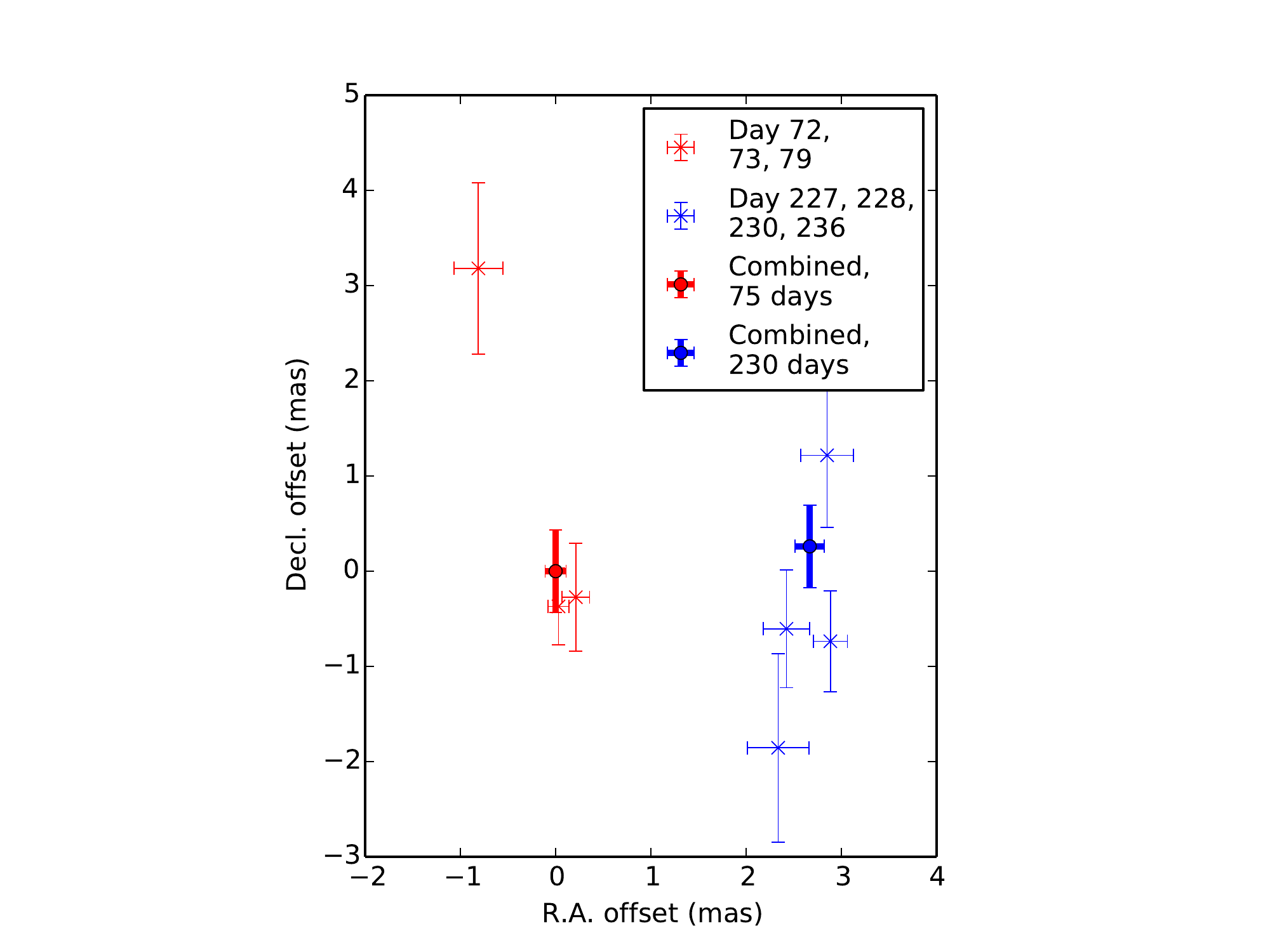} &
\includegraphics[trim=4.5cm 0cm 4.5cm 0cm, width=3in, clip]{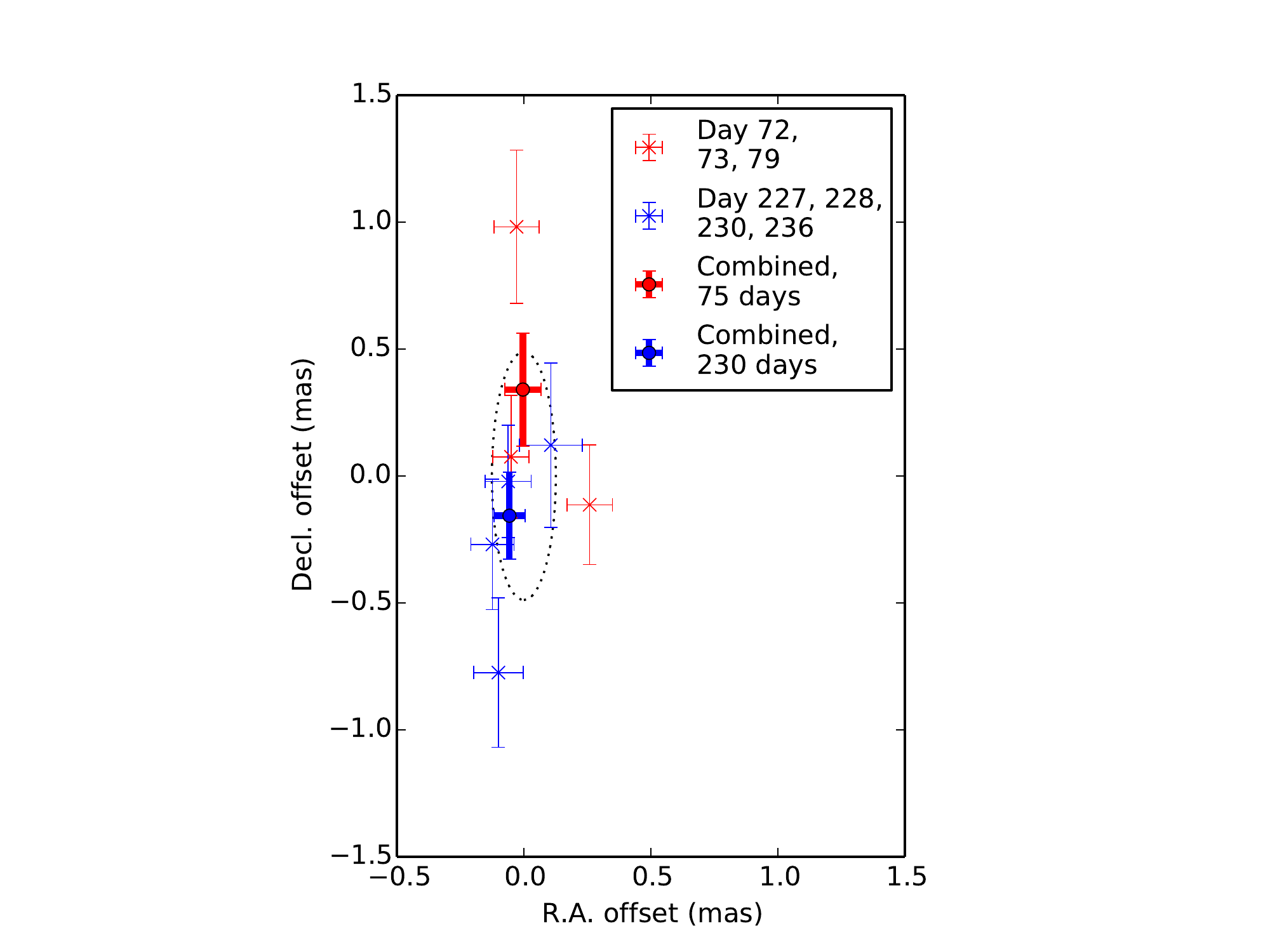}
\end{tabular}
\caption{{\bf VLBI astrometric accuracy.}
The VLBI positions of GW170817 (panel a, relative to the best-fit position at day 75)  and the low luminosity AGN in NGC 4993 (panel b, relative to the previously derived position using VLBA-only observations). The individual observations of GW170817 have very low S/N and hence large errors; the moderately discrepant measurement on day 72 has the lowest S/N and was affected by observing issues at the Green Bank Telescope. The NGC 4993 positions do not show any significant systematic position shifts between the two epochs, and are consistent with our estimated systematic position uncertainties of 0.15 mas in right ascension and 0.5 mas in declination. All errorbars/uncertainties quoted are 1$\sigma$.}
\label{fig:position-check}
\end{figure}

%%
%% TABLES
%%
%% If there are any tables, put them here.
%%

\end{document}